\documentclass{article}

\usepackage{amssymb,amsfonts,amsmath,stmaryrd}
\usepackage{cite,enumerate,float,indentfirst}
\usepackage{color}

\def\be{\begin{eqnarray}}
\def\ee{\end{eqnarray}}
\def\nn{\nonumber}

\def\tr{{\rm tr}\,}

\definecolor{red}{rgb}{1,0,0}
\definecolor{orange}{rgb}{1,0.5,0}
\definecolor{violet}{rgb}{0.7,0,1}

\hoffset= - 1.0in         

\begin{document}

\hfill
ITEP/TH-36/18

\hfill

\bigskip

\centerline{\Large{
 Cauchy formula and the character ring
}}

\bigskip

\centerline{{\bf A.Morozov}}

\bigskip

\centerline{\it ITEP \& IITP, Moscow, Russia}

\bigskip

\centerline{ABSTRACT}

\bigskip

{\footnotesize
Cauchy summation formula plays a central role in application
of character calculus to many problems, from AGT-implied
Nekrasov decomposition of conformal blocks to topological-vertex
decompositions of link invariants.
We briefly review the equivalence between Cauchy formula and
expressibility of skew characters through the Littlewood-Richardson
coefficients.
As  not-quite-a-trivial illustration we consider how this
equivalence works in the case of plane partitions --
at the simplest truly interesting level of just  four boxes.
}

\bigskip

\bigskip

\section{ Introduction}

As anticipated long ago \cite{UFN3},
Schur functions and their various generalizations,
like Macdonald \cite{Mac} and Kerov \cite{Kerov} functions,
generalized Macdonald polynomials \cite{genMac},
tensor-model characters \cite{tenmodchar} and the still-hypothetical
3-Schur functions \cite{3Schurs} play an increasing role in modern theory,
especially in consideration of essentially non-perturbative phenomena.
Technically they appear at least in three different contexts:
\begin{itemize}

\item{in formal representation theory --
as characters of $Sl_N$ representations $S_R[X]={\rm Tr}_R {\cal X}^{(R)}$,
and thus as the building blocks for integrable tau-functions
through the general construction, reviewed in \cite{GKLMM}
}

\item{in decomposition formulas for the integrands of
free-field screening correlators like
\be
\Big<\prod_i e^{\phi(x_i)} \prod_j e^{-\phi(y_j)}\Big>_{\phi} \sim
\prod_{i,j} (x_i-y_j)^{-1}  \sim \sum_R S_R[X]S_{R^\vee}[Y^{-1}]
\label{ffcorr}
\ee
}
\item{and as preserved quantities in Selberg-Kadell-type integrals \cite{Kad},
which stand behind the basic {\it super}\-integra\-bility/localization
property \cite{MMchar}
\be
\Big< S_R[X] \Big>_X \sim S_R[X^*]
\label{charchar}
\ee
actually serving as a selection rule for "good" theories,
which provide "matrix-model $\tau$-functions" \cite{UFN3}
after (functional) integration over fields.
}
\end{itemize}

One of the many widely-known  examples of (\ref{charchar}),
appears when we integrate {\it exactly} over $x$ and $y$ in (\ref{ffcorr})
to obtain a correlator of screenings $\oint e^{\phi(x)} dx$ \cite{Kad}.
In this case the a combination of (\ref{ffcorr}) and (\ref{charchar}) immediately
provides the AGT-induced \cite{AGT} Nekrasov decomposition \cite{Nek} of conformal blocks,
realized in terms of  conformal (Dotsenko-Fateev) matrix models \cite{confmamo}
-- and their far-going network-model generalizations \cite{network}.
Further steps on this way, as well as a related development with
matrix $\longrightarrow$ tensor model generalizations require essential
extension of the theory of Schur characters in various directions.
Some things, however, should supposedly remain intact -- and serve as the
carrying construction of this future general theory.

In this short note we consider two of such properties:
Cauchy  decomposition formula which stands behind (\ref{ffcorr})
and the  skew-character decomposition, which plays the central role
in technical applications of Schur functions to representation theory.
These two properties are in fact intimately related: imposing one implies another.
This is a simple but important remark,
it considerably {\it weakens} their impact on generalizations:
one restriction (to keep these properties) is much less than two.
Also, it reduces the number of "miracles" and thus the attractiveness of particular
generalization attempts.
After a brief presentation of the formal relation
we present an explicit example -- of the problems with the building project
of the 3-Schur functions, encountered at the level of the size-four plain partitions,
where the (general) relation between Cauchy and skew decompositions shows up in a somewhat
unusual way.

\section{ Cauchy vs skew
\label{Cvskew}}
Imagine that we have a set of functions $S_\sigma\{ p\}$ which depend
on a multi-component $p_k$, $k\in K$ and are labeled by  elements $\sigma\in \Sigma$
of some set $\Sigma$.
Let them form a full linear basis in the space of functions of $\{p\}$.
Then they also form a closed ring under the ordinary multiplication
\be
S_{\sigma'}\{ p\}\cdot S_{\sigma''} \{p\} =
\sum_{\sigma\in\Sigma}N_{\sigma'\sigma''}^\sigma S_\sigma\{p\}
\label{ring}
\ee
with some structure constants $N$ (not obligatory integer).
In this setting there is an obvious
equivalence between two different-looking of statements:
Cauchy summation formula and decomposition rule of the skew-functions.

Cauchy formula states that
\be
\boxed{
\sum_{\sigma\in \Sigma} \frac{S_\sigma\{  p\} S_\sigma\{p'\}}{||S_\sigma||^2}
= \exp\left(\sum_{k\in K} \frac{  p_k   p'_k}{||p_k||^2}\right)
}
\label{Cauchy}
\ee
with a certain norm in  the space of $\{p\}$-variables.
Ideally one can think of a scalar product, with respect to which
both $p_k$ and $S_\sigma\{p\}$ are orthogonal:
\be
\Big<p_k\Big|p_k'\Big> = ||p_k||^2\cdot \delta_{k,k'}
\ee
\be
\Big<S_\sigma\{p\}\Big|S_{\sigma'}\{p\}\Big>
= ||S_\sigma||^2\cdot \delta_{\sigma,\sigma'}
\label{Sprod}
\ee
However,  really important is the {\bf bilinear exponent}.
As a corollary, by multiplying two copies of (\ref{Cauchy})
with the same $p$ but different $p'$, we get:
\be
\left(\sum_{\sigma'\in \Sigma}
\frac{S_{\sigma'}\{  p\} S_{\sigma'}\{p'\}}{||S_{\sigma'}||^2}\right)\cdot
\left(\sum_{\sigma''\in \Sigma}
\frac{S_{\sigma''}\{  p\} S_{\sigma''}\{p''\}}{||S_{\sigma''}||^2}\right)
&\stackrel{(\ref{Cauchy})}{=}
& \exp\left(\sum_{k\in K} \frac{  p_k   (p'_k+p_k'')}{||p_k||^2}\right)
\nn \\ \nn \\
|| \ (\ref{ring}) \ \ \ \ \ \ \ \ \ \ \ \ \ \ \ \ \
&& \ \ \ \ \ \ \ \ \ \ \ \ \ \ \ \ \ || \ (\ref{Cauchy})
\nn \\ \nn \\
\sum_{\sigma',\sigma'',\sigma\in \Sigma} N_{\sigma'\sigma''}^\sigma
S_\sigma\{p\}\cdot\frac{S_{\sigma'}\{p'\}}{||S_{\sigma'}||^2}
\cdot\frac{S_{\sigma''}\{p''\}}{||S_{\sigma''}||^2}
&=&
\sum_\sigma S_\sigma\{p\}\cdot \frac{S_\sigma\{p'+p''\}}{||S_\sigma||^2}
\label{Cauchysum}
\ee
If we now consider the function of $p'+p''$ as that of $p'$, we obtain
\be
S_\sigma\{p'+p''\}  = \sum_{\sigma'\in\Sigma}
S_{\sigma/\sigma'}\{p''\}\cdot  S_{\sigma'}\{p'\}
\label{skewdef}
\ee
where the $p''$-dependent coefficients are known as {\it skew}-functions.
Then equivalence of the two relations in the second line of (\ref{Cauchysum})
implies that
\vspace{-0.4cm}
\be
\boxed{
S_{\sigma/\sigma'}\{p''\}
=  \sum_{\sigma''\in\Sigma} {\bf N}^\sigma_{\sigma'\sigma''}
\cdot  S_{\sigma''}\{p''\}
}
=  \sum_{\sigma''\in\Sigma}
\overbrace{\frac{||S_\sigma||^2}{||S_{\sigma'}||^2||S_{\sigma''}||^2}
N^\sigma_{\sigma'\sigma''} }^{{\bf N}^\sigma_{\sigma'\sigma''}}
\cdot  S_{\sigma''}\{p''\}
\label{decoskew}
\ee
with the same structure constants $N$ as in (\ref{ring}).
The differently-normalized bold-faced ${\bf N}$ are instead the structure
constants in multiplication of "dual" functions (boldfaced):
\be
{\bf S}_{\sigma' }\{ p\}\cdot{\bf S}_{\sigma''} \{p\} :=
\frac{S_{\sigma'}\{ p\}}{||S_{\sigma'}||^2}\cdot \frac{S_{\sigma''}
\{p\}}{||S_{\sigma''}||^2} =
\sum_{\sigma\in\Sigma}  {\bf N}_{\sigma',\sigma''}^\sigma
\frac{S_\sigma\{p\}}{||S_\sigma||^2}
= \sum_{\sigma\in\Sigma}  {\bf N}_{\sigma',\sigma''}^\sigma {\bf S}_\sigma\{p\}
\label{ringdu}
\ee
Thus we see that (\ref{Cauchy}) implies (\ref{decoskew}).

This statement can be partly inverted: if the skew functions in (\ref{skewdef})
possess the expansion (\ref{decoskew}) with the same structure constants
as in (\ref{ring}), this implies {\it some} version of Cauchy summation
formula (\ref{Cauchy}) with a bilinear exponent --
but, strictly speaking, with some unspecified coefficients  at the place of $||p_k||^{-2}$.

To avoid possible confusion, (\ref{ring}) and (\ref{ringdu})
are {\it not} the {\it statements} --
these are just the {\it definitions} of the structure constants $N$ and ${\bf N}$
for a given set of functions $S_\sigma\{p\}$.
Of course, one can instead use (\ref{decoskew}) as a definition of ${\bar N}$,
like it was done in \cite{3Schurs}, -- then
the statement will be that  (\ref{ringdu})
depends on  validity of some version of (\ref{Cauchy}).

\section{Particular cases}

So far, in most applications in physics the set $\Sigma$ is that of Young diagrams
(partitions of integers) --
this is especially natural for applications to representations
of linear and symmetric groups $Gl_N$ and ${\cal S}_N$.
Then the relevant set $K$ is just that of natural numbers:
the "time variables" are just $\{p_1,p_2,\ldots\}$, and these are exactly
enough to "enumerate" all Young diagrams by the rule
\be
R= [r_1\geq r_2\geq \ldots \geq r_{l_R}>0] = [l_R^{m_{l_R}},\ldots,3^{m_3},2^{m_2},1^{m_1} ]
\ \ \longleftrightarrow \ \
p^R = \prod_{k=1}^{l_R} p_k^{m_k} = \prod_{a=1}^{l_R} p_{r_a}
\ee
Relation to representation theory and conformal matrix/network models
in (\ref{ffcorr}) and (\ref{charchar}),
appears on the Miwa locus $p_k=\tr X^k$ with the $N\times N$ matrix $X$,
which in representation $R$ becomes a matrix ${\cal X}^{(R)}$ of the size
${\rm dim}_R = {\rm Schur}_R[I] = {\rm Schur}_R\{p_k=N\}$,
made from the $N$ eigenvalues of $X$.
Associated scalar product is usually taken to be
\vspace{-0.4cm}
\be
\Big< p^R\Big|p^{R'}\Big>^{(g)} =
\delta_{R,R'}\cdot \overbrace{\prod_{a=1}^{l_R} a^{m_a}\, m_a!}^{z_R}\, g_a^{m_a}
\label{scapr}
\ee
For all $g_a=1$ we get the Schur functions {\it per se},
the corresponding factor $z_R$ is the one which appears in the orthogonality
condition for symmetric-group characters $\psi_R(\Delta)$,
\be
\sum_{\Delta\vdash |R|} \frac{\psi_R(\Delta)\psi_{R'}(\Delta)}{z_\Delta} = \delta_{R,R'}
\ \ \ \ \ \ \Longleftrightarrow \ \ \ \ \ \ \
\sum_{R\vdash \Delta}  \psi_R(\Delta)\psi_R(\Delta')  = z_\Delta\,\delta_{\Delta,\Delta'}
\ee
and the structure constants $N^R_{R'R''}$ in (\ref{ring}) are  the integer-valued
Richardson-Littlewood coefficients, counting multiplicities of representation $R$
in the product $R'\otimes R''$.
In deformation to Macdonald polynomials, when
\be
g_a = \frac{q^a-q^{-a}}{t^a-t^{-a}}
\label{macloc}
\ee
these $N$ become functions  of $q$  and $t$, still they vanish whenever
$R\notin R'\otimes R''$.

For arbitrary parameters $g_a$ we get Kerov functions \cite{Kerov},
for them  the restriction on  $R$ is softened
to a one, natural for the Young-diagrams {\it per se}:
\be
N^R_{R'R''} \neq 0 \ \ \ \ \Longrightarrow \ \ \ \
R'+R'' \leq R \leq R\cup R'
\label{sumdia}
\ee
and exact relation to representation theory of $SL_\infty$ and ${\cal S}_\infty$
is lost.
Still the absolute majority of other properties, including the Cauchy and
skew-Kerov decompositions remain true --
and application of generic Kerov functions to physical theories is just a matter of time
(see \cite{Kerovapp} for the first examples).

However, already for the by-now-conventional applications,
restriction to $\Sigma=\{partitions\}$ is insufficient.
Nekrasov calculus for generic $\Omega$-backgrounds
(for $c\neq 1$, i.e. $\epsilon_1\neq -\epsilon_2$) requires "generalized"
Macdonald functions \cite{genMac}, depending on collections (strings) of Young diagrams.
This, however, is not a very big problem -- it is enough just to consider several copies
of time variables, though the scalar product can require non-trivial modification
\cite{MMgenmac}.
More challenging are the
ordered sequences of Young diagrams (forming the plane partitions),
which are needed in generic network models and representation theory of DIM-algebras.
The corresponding "triple-Macdonald polynomials", though constructible
in terms of the ordinary ones \cite{Zen}, should depend on a very different
set $K$ of time-variables and be described by a more-first-principle theory.

One of the fresh related directions is the basic tangles-calculus \cite{tangles} relation
\be
H^{\rm Hopf}_{(R'\otimes R'')\times Q} =
H^{\rm Hopf}_{R'\times Q} \cdot H^{\rm Hopf}_{R''\times Q}
\ee
for the properly normalized colored Hopf-link invariants,
which provides for them an interpretation as $Q$-dependent characters
(note that this is a manifestation of the rule (\ref{charchar}),
because these invariants are averages of Wilson loops
${\rm Tr}_R P\exp\left(\oint{\cal A}\right)$, which are themselves the
{\it gauge-field}-dependent
characters in Chern-Simons theory).
Since Hopf invariants are supposedly related to topological vertices \cite{TV}
(DIM-algebra intertwiners),
this has direct connection to the still-underdeveloped representation theory of DIM
algebras.

An ever further-going challenge is adequate description of tensor-model characters,
where some "non-abelization" looks unavoidable already at the level of
(\ref{charchar}) -- straightforward lifting of Schur functions to these theories
does not seem to provide a full basis in the operator space \cite{tenmodchar}.
In this note we do not go as far as full-fledged tensor-model considerations,
but provide just a simple example of difficulties, encountered at the plain-partition stage.
We demonstrate that, conversely to possible expectations, Cauchy formula is considerably
easier to satisfy than building a true collection of 3-Schur functions.

\section{The 3-Schur attempt}

When we switch from the ordinary to plane partitions in the role of the set $\Sigma$,
the first thing to change is the set $K$ of time-variables.
In order for the space polynomials of $p_k$ to have the same dimension as that
of the plane partitions we need $p_{k,i}$ with integer $1\leq i\leq k$ with the grading
degree $\sum_{i\leq k} kp_{k,i}$.
Then at the "level" (degree) one we have just a single monomial $p_{1,1}$ and a single
plane partition with one box,
at level two -- three monomials $p_{2,1}$, $p_{2,2}$, $p_{1,1}^2$  and three plane
partitions with two boxes and so on.
Since the grading does not depend on $i$ it can be convenient to speak of the $k$
dimensional vector spaces and denote the time variables $\vec p_k$ --
assuming that the number of vector components is $k$.
The 3-Schur functions should be homogeneous functions of these variables and
form a full basis -- and thus a ring.
However, the first naive attempt in \cite{3Schurs} to build these functions
runs into problems, which we will now try to illustrate.
This attempt was build on two postulates: that the scalar product does not depend on $i$
and is given by the same formula (\ref{scapr}) with all $g_k=1$,
\be
\Big< \prod_{i\leq k} p_{k,i}^{m_{k,i}}\Big|\prod_{i\leq k} p_{k,i}^{m'_{k,i}}\Big>^{(g)} =
\prod_{i\leq k}\delta_{m_{k,i},m'_{k,i}}
\cdot k^{m_{k,i}}\, m_{k,i}! \, g_k^{m_{k,i}}
\label{scaprmod}
\ee
and that the multiplication operation (\ref{ring}) is dictated by "natural" composition
of plane partitions, see below.
Both postulates are not very well justified, but it is instructive to see what is
exactly the problem they lead to.

We denote the three dimensions of the space where the plane partitions lie, by $x,y,z$
and use $\rho=\{x,y,z\}$ as a label.
When the number of boxes is small, partitions  lie entirely in one of the three planes
and can be labeled by Young diagrams together with the ordered pair of indices $x,y,z$.
When there is just one column/row, only one index remains.
For symmetric Young diagrams the order does not matter and it is also convenient to
use orthogonal direction $z$ instead of $xy\cong yx$.
Then the  3-Schur functions at the first three levels are:
\be
{\cal S}_{[1]} = p_1  \ \ \ \ \ \ \ \ \ \ \
&{\cal S}_{[2]}^\rho = \frac{\vec\alpha_2^\rho \vec p_2 + p_1^2}{2}  \ \ \ \ \ \ \ \ \ \ \
& \begin{array}{c}
{\cal S}_{[3]}^\rho = \frac{\vec\alpha_3^\rho \vec p_3}{3}
+ \frac{\vec\alpha_2^\rho\vec p_2\,p_1}{2} + \frac{p_1^3}{6}
\\ \\
{\cal S}_{[2,1]}^\rho = \frac{\vec\beta_3^\rho \vec p_3}{3}
- \frac{\vec\alpha_2^\rho\vec p_2\,p_1}{2} + \frac{p_1^3}{3}
\end{array}
\label{3Schurs123}
\ee
The have simple $\rho$-independent norms
$ \ \ \ \
||{\cal S}_{[1]}||^2=1, \ \ \    ||{\cal S}_{[2]}||^2 = \frac{3}{2}, \ \ \
||{\cal S}_{[3]}||^2 = \frac{9}{2}, \ \ \ \ \
||{\cal S}_{[2,1 ]}||^2 = \frac{9}{4} \ \ \
$
and  satisfy the relation (\ref{ringdu}) and (\ref{decoskew}) in the most natural way:
\be
{\cal S}_{[1]}^2 =  \frac{\sum_{\rho=x,y,z}{\cal S}^\rho_{[2]}}{||S_{[2]}||^2}
\ \ \ \Longleftrightarrow \ \ \
\Delta {\cal S}_{[2]}^\rho:= {\cal S}_{[2]}^\rho\{p'+p''\}
-{\cal S}_{[2]}^\rho\{p'\}-{\cal S}_{[2]}^\rho\{p''\}
= S_{[1]}^\rho\{p'\}S_{[1]}^\rho\{p''\} := S_{[1]}^\rho\otimes S_{[1]}^\rho
\ee
\be
\frac{{\cal S}_{[2]}^\rho\cdot {\cal S}_{[1]}}{||{\cal S}_{[2]}||^2}
= \frac{ {\cal S}^\rho_{[3]}}{||S_{[3]}||^2}
+ \frac{\sum_{\rho'\neq \rho} {\cal S}^{\rho'}_{[2,1]}}{||S_{[2,1]}||^2}
&\! \Longleftrightarrow   & \!\!\!
\begin{array}{c}
\Delta {\cal S}_{[3]}^\rho = {\cal S}_{[2]}^\rho\otimes {\cal S}_{[1]}
+ {\cal S}_{[1]}\otimes {\cal S}_{[2]}^\rho
\\ \\
\Delta {\cal S}_{[2,1]}^\rho
= \Big({\cal S}_{[2]}^{\rho'}+{\cal S}_{[2]}^{\rho''}\Big)\otimes {\cal S}_{[1]}
+ {\cal S}_{[1]}\otimes \Big({\cal S}_{[2]}^{\rho'}+{\cal S}_{[2]}^{\rho''}\Big)
\end{array}
\label{lev3rels}
\ee
Despite we put here the sign $ \Longleftrightarrow $ we know from sec.\ref{Cvskew}
that such an identical correspondence between multiplication  and decomposition
should be tied to validity of Cauchy formula --
and indeed it is true:
\be
1+ {\cal S}_{[1]}\{p\}{\cal S}_{[1]}\{p'\}
+\sum_{\rho=x,y,z}\frac{{\cal S}_{[2]}^\rho\{p\}{\cal S}_{[2]}^\rho\{p'\}}{||S_{[2]}||^2}
+\sum_{\rho=x,y,z}\frac{{\cal S}_{[3]}^\rho\{p\}{\cal S}_{[3]}^\rho\{p'\}}{||S_{[3]}||^2}
+\sum_{\rho=x,y,z}\frac{{\cal S}_{[2,1]}^\rho\{p\}{\cal
S}_{[2,1]}^\rho\{p'\}}{||S_{[2,1]}||^2}
= \nn \\
= 1 +p_1p_1' + \frac{\vec p_2\vec p_2  \,\!\!'+(p_1p_1')^2}{2}
+ \frac{\vec p_3\vec p_3 \,\!\!'}{3}+\frac{(\vec p_2\vec p_2 \,\!\!')(p_1p_1')}{2}
+  \frac{(p_1p_1')^3}{6}
\ \ \ \ \ \ \ \
\ee
Moreover, in accordance with the still another {\it natural expectation}
(\ref{Sprod}), all these ${\cal S}$-functions are mutually orthogonal
\be
\Big<{\cal S}^\rho_{[2]}\Big| {\cal S}^{\rho'}_{[2]}\Big>
= ||{\cal S}_{[2]}||^2\,\delta^{\rho,\rho'}, \ \ \ \
\Big<{\cal S}^\rho_{[3]}\Big| {\cal S}^{\rho'}_{[3]}\Big>
= ||{\cal S}_{[3]}||^2\,\delta^{\rho,\rho'}, \ \
\Big<{\cal S}^\rho_{[2,1]}\Big| {\cal S}^{\rho'}_{[2,1]}\Big>
= ||{\cal S}_{[2,1]}||^2\,\delta^{\rho,\rho'}, \ \
\Big<{\cal S}^\rho_{[3]}\Big| {\cal S}^{\rho'}_{[2,1]}\Big> = 0
\ee
To check all these formulas one needs to substitute explicit expressions for the
Mercedes-star
vectors \cite{3Schurs}:
{\footnotesize
\be
\vec\alpha_2^x = \Big(-\frac{1}{\sqrt{2}},\sqrt{\frac{3}{2}}\Big),   \ \ \ \
\vec\alpha_2^y = \Big(-\frac{1}{\sqrt{2}},-\sqrt{\frac{3}{2}}\Big),  \ \ \ \
\vec\alpha_2^z = (\sqrt{2},0)
\nn \\
\vec\alpha_3^x = \Big(- \sqrt{\frac{3}{2}},\frac{3}{\sqrt{2}},-2\Big)
=-2(\beta_3^y+\beta_3^z),   \ \ \ \
\vec\alpha_3^y = \Big(-\sqrt{\frac{3}{2}},-\frac{3}{\sqrt{2}},-2\Big)
=-2(\beta_3^y+\beta_3^z),  \ \ \ \
\vec\alpha_3^z = (\sqrt{6},0,-2)=-2(\beta_3^x+\beta_3^y)
\nn \\
 \vec\beta_3^x = \Big(- \sqrt{\frac{3}{8}},\frac{3}{\sqrt{8}},-\frac{1}{2}\Big),   \ \ \ \
\vec\beta_3^y = \Big(- \sqrt{\frac{3}{8}},-\frac{3}{\sqrt{8}},-\frac{1}{2}\Big),  \ \ \ \
\vec\beta_3^z = \Big( \sqrt{\frac{3}{2}},0,-\frac{1}{2}\Big)
\nn
\ee
}
Note that the relation
$\vec\alpha^\rho_3 ||{\cal S}_{[3]}||^{-2}
+(\vec\beta^{\rho'}_3+\beta^{\rho''}_{[3]})||{\cal S}_{[2,1]}||^{-2}=0$
between $\vec\alpha_3$ and $\vec\beta_3$
is necessary for the l.h.s. of (\ref{lev3rels}) to hold,
because ${\cal S}_{[2]}^\rho{\cal S}_{[1]}$
there does not depend on ${\vec p_3}$.

\section{Expectation at level four}

The first truly interesting level is four, when one of $13$ plane partitions,
which we denote by $\Yup$, is essentially 3-dimensional.
The "natural" multiplication and decomposition rules in this case seem to be
\be
\frac{{\cal S}_{[3]}^\rho\cdot {\cal S}_{[1]}}{||{\cal S}_{[3]}||^2}
\ \stackrel{?}{=}\ \frac{{\cal S}_{[4]}^\rho }{||{\cal S}_{[4]}||^2}
+\frac{ {\cal S}_{[3,1]}^{\rho\rho'}+ {\cal S}^{\rho\rho''}_{[3,1]}}
{||{\cal S}_{[3,1]}||^2}
\nn \\
\frac{{\cal S}_{[2,1]}^\rho\cdot {\cal S}_{[1]}}{||{\cal S}_{[2,1]}||^2}
\ \stackrel{?}{=}\  \frac{ {\cal S}_{[3,1]}^{\rho'\rho''}+ {\cal S}^{\rho''\rho'}_{[3,1]}}
{||{\cal S}_{[3,1]}||^2}
+ \frac{ {\cal S}_{[2,2]}^\rho}{||{\cal S}_{[2,2]}||^2}
+  \frac{ {\cal S}_{_\Yup}  }{||{\cal S}_{_\Yup}||^2}
\nn \\
\frac{{\cal S}_{[2]}^{\rho'}\cdot {\cal S}_{[2]}^{\rho''} }{||{\cal S}_{[2]}||^4}
\ \stackrel{?}{=}\
\frac{ {\cal S}_{[3,1]}^{\rho'\rho''}+ {\cal S}^{\rho''\rho'}_{[3,1]}}
{||{\cal S}_{[3,1]}||^2}
+ \frac{ {\cal S}_{_\Yup}  }{||{\cal S}_{_\Yup}||^2}
\nn\\
 \frac{({\cal S}_{[2]}^\rho)^2}{||{\cal S}_{[2]}||^4}
\ \stackrel{?}{=}\  \frac{{\cal S}_{[4]}^\rho }{||{\cal S}_{[4]}||^2}
+\frac{ {\cal S}_{[3,1]}^{\rho\rho'}+ {\cal S}^{\rho\rho''}_{[3,1]}}
{||{\cal S}_{[3,1]}||^2}
+ \frac{ {\cal S}_{[2,2]}^{\rho'}+ {\cal S}_{[2,2]}^{\rho''}}{||{\cal S}_{[2,2]}||^2}
\label{multilev4}
\ee
and, "accordingly",
\be
\Delta {\cal S}^\rho_4 =
\Big({\cal S}^{\rho\rho'}_{[3,1]}+{\cal S}^{\rho\rho''}_{[3,1]}\Big)\otimes S_{[1]}
+ S_{[1]}\otimes \Big({\cal S}^{\rho\rho'}_{[3,1]}+{\cal S}^{\rho\rho''}_{[3,1]}\Big)
+ {\cal S}^\rho_{[2]}\otimes {\cal S}^{\rho}_{[2]}
\nn \\
\Delta {\cal S}^{\rho\rho'}_{[3,1]} =
({\cal S}^\rho_{[3]}+{\cal S}^{\rho''}_{[2,1]})\otimes {\cal S}_{[1]}
+{\cal S}^\rho_{[2]}\otimes{\cal S}^{\rho}_{[2]}
+{\cal S}^\rho_{[2]}\otimes{\cal S}^{\rho'}_{[2]}
+{\cal S}^{\rho'}_{[2]}\otimes{\cal S}^{\rho}_{[2]}
+ {\cal S}_{[1]}\otimes ({\cal S}^\rho_{[3]}+{\cal S}^{\rho''}_{[2,1]})
\nn \\
\Delta {\cal S}^{\rho }_{[2,2]} = {\cal S}_{[2,1]}^\rho\otimes {\cal S}_{[1]}
+ {\cal S}^{\rho'}_{[2]}\otimes{\cal S}^{\rho'}_{[2]}
+{\cal S}^{\rho''}_{[2]}\otimes{\cal S}^{\rho''}_{[2]}
+{\cal S}_{[1]}\otimes{\cal S}_{[2,1]}^\rho
\nn \\
\Delta {\cal S}_{_\Yup} =
\sum_\rho \Big({\cal S}_{[2,1]}^\rho\otimes{\cal S}_{[1]}
+ {\cal S}_{[1]}\otimes {\cal S}_{[2,1]}^\rho\Big)
+ \sum_{\rho\neq\rho'} {\cal S}^\rho_{[2]}\otimes {\cal S}^{\rho'}_{[2]}
\label{decolev4}
\ee
If true, these (\ref{decolev4}) would imply that
\vspace{-0.5cm}
\be
{\cal S}^\rho_{[4]} = \frac{\vec\alpha_4^\rho\vec p_4}{4}
+ \overbrace{\frac{\vec\alpha_3^\rho\vec p_3\, p_1}{3}
+ \frac{(\vec\alpha_2^\rho\vec p_2)^2}{8}
+ \frac{\vec\alpha_2^\rho\vec p_2 p_1^2}{4} + \frac{p_1^4}{24}}^{\tilde{\cal S}_{[4]}}
\nn \\
{\cal S}^{\rho\rho'}_{[3,1]} =
\frac{\vec\beta_4^{\rho\rho'}\vec p_4}{4}
+ \overbrace{\frac{(\vec\alpha_3^{\rho}+\beta_3^{\rho''})\vec p_3\, p_1}{3}
+ \frac{(\vec\alpha_2^\rho\vec p_2)^2}{8}
+ \frac{(\vec\alpha_2^{\rho}\vec p_2)(\vec\alpha_2^{\rho'}\vec p_2)}{4}
+ \frac{(2\vec\alpha_2^\rho+\vec\alpha_2^{\rho'})\vec p_2 p_1^2}{4}
+ \frac{p_1^4}{8}}^{\tilde {\cal S}^{\rho\rho'}_{[3,1]}}
\nn \\
{\cal S}^\rho_{[2,2]} = \frac{\vec\gamma_4^\rho\vec p_4}{4}
+ \overbrace{\frac{\vec\beta_3^\rho\vec p_3\, p_1}{3}
+  \frac{(\vec\alpha_2^\rho\vec p_2)^2}{8}
- \frac{(\vec\alpha_2^{\rho'}\vec p_2)(\vec\alpha_2^{\rho''}\vec p_2)}{4}
- \frac{\vec\alpha_2^\rho\vec p_2 p_1^2}{4}
+ \frac{p_1^4}{12}}^{\tilde {\cal S}^\rho_{[2,2]}  }
\nn \\
{\cal S}_\Yup = \frac{\vec\delta_4^\rho\vec p_4}{4}
+ \overbrace{\sum_{\rho=x,y,z}\left(\frac{\vec\beta_3^\rho\vec p_3\, p_1}{3}
- \frac{(\vec\alpha_2^\rho\vec p_2)^2}{8}\right)
  + \frac{p_1^4}{4}}^{\tilde {\cal S}_\Yup  }
\ee
As usual, only the $\vec p_4$-independent parts of these formulas,
which we denote by tildes, are prescribed by (\ref{decolev4}).
Likewise, only these pieces are seen in multiplication formulas --
irrespective of their exact shape and relation to decompositions (\ref{decolev4}),
i.e. irrespective of the literal validity of (\ref{multilev4}).

If expectation of \cite{3Schurs} was fully correct,
both (\ref{multilev4}) and (\ref{decolev4}) would hold --
what, as we know, imply also the validity of Cauchy formula
\be
\sum_{\rho=x,y,z}\frac{{\cal S}_{[4]}^\rho\{p\}{\cal S}_{[4]}^\rho\{p'\}}{||S_{[4]}||^2}
+\sum_{\rho\neq\rho' }\frac{{\cal S}_{[3,1]}^{\rho\rho'}\{p\}
{\cal S}_{[3,1]}^{\rho\rho'}\{p'\}}{||S_{[3,1]}||^2}
+\sum_{\rho=x,y,z}\frac{{\cal S}_{[2,2]}^\rho\{p\}{\cal
S}_{[2,2]}^\rho\{p'\}}{||S_{[2,2]}||^2}
+ \frac{{\cal S}_{_\Yup} \{p\}{\cal S}_{_\Yup} \{p'\}}{||S_{_\Yup}||^2}
\stackrel{?}{=} \nn \\
\stackrel{?}{=}
\frac{\vec p_4\vec p_4 \,\!\!'}{4}
+\frac{(\vec p_3\vec p_3 \,\!\!')(p_1p_1')}{3}
+ \frac{(\vec p_2\vec p_2 \,\!\!')^2}{8}
+\frac{(\vec p_2\vec p_2 \,\!\!')(p_1p_1')^2}{4} +  \frac{(p_1p_1')^4}{24}
\label{Cauchylev4}
\ee
and, in the dream case, also orthogonality conditions
\be
\Big<{\cal S}^\rho_{[4]}\Big| {\cal S}^{\rho'}_{[4]}\Big>
\stackrel{?}{=} ||{\cal S}_{[4]}||^2\,\delta^{\rho,\rho'}, \ \ \ \
\Big<{\cal S}^{\rho\rho'}_{[3,1]}\Big| {\cal S}^{\rho''\rho'''}_{[3]}\Big>
\stackrel{?}{=} ||{\cal S}_{[3,1]}||^2\,\delta^{\rho,\rho''}\delta^{\rho',\rho'''}, \ \
\Big<{\cal S}^\rho_{[2,2]}\Big| {\cal S}^{\rho'}_{[2,2]}\Big>
\stackrel{?}{=} ||{\cal S}_{[2,2]}||^2\,\delta^{\rho,\rho'}, \nn \\
\Big<{\cal S}^\rho_{[4]}\Big| {\cal S}^{\rho'\rho''}_{[3,1]}\Big> \stackrel{?}{=}
\Big<{\cal S}^\rho_{[4]}\Big| {\cal S}^{\rho'}_{[2,2]}\Big> \stackrel{?}{=}
\Big<{\cal S}^\rho_{[4]}\Big| {\cal S}_{_\Yup}\Big> \stackrel{?}{=}
\Big<{\cal S}^{\rho\rho'}_{[3,1]}\Big| {\cal S}^{\rho''}_{[2,2]}\Big> \stackrel{?}{=}
\Big<{\cal S}^{\rho\rho'}_{[3,1]}\Big| {\cal S}^{\rho'}_{_\Yup}\Big> \stackrel{?}{=}
\Big<{\cal S}^\rho_{[2,2]}\Big| {\cal S}_{_\Yup}\Big> \stackrel{?}{=} 0
\ee

\section{The situation   at level four}

Given explicit expressions (\ref{3Schurs123}) we can check the $\vec p_4$-independent
parts of (\ref{multilev4}) and (\ref{Cauchylev4}).
It turns out that instead of them we have similar, still very different relations:
\be
\begin{array}{c|c}
\text{expected} & \text{true}\\
\hline
& \\
\frac{{\cal S}_{[3]}^\rho\cdot {\cal S}_{[1]}}{||{\cal S}_{[3]}||^2}
\ \stackrel{?}{=}\ \boxed{\frac{ \tilde{\cal S}_{[4]}^\rho }{||{\cal S}_{[4]}||^2}}
+\frac{\tilde {\cal S}_{[3,1]}^{\rho\rho'}+\tilde {\cal S}^{\rho\rho''}_{[3,1]}}
{||{\cal S}_{[3,1]}||^2}
&
\frac{{\cal S}_{[3]}^\rho\cdot {\cal S}_{[1]}}{||{\cal S}_{[3]}||^2}
= \frac{\tilde{\cal S}_{[3,1]}^{\rho\rho'}+\tilde{\cal
S}^{\rho\rho''}_{[3,1]}}{\frac{27}{4}}
\nn \\ & \\
\frac{{\cal S}_{[2,1]}^\rho\cdot {\cal S}_{[1]}}{||{\cal S}_{[2,1]}||^2}
\ \stackrel{?}{=}\ \frac{ \tilde{\cal S}_{[3,1]}^{\rho'\rho''}
+ \tilde{\cal S}^{\rho''\rho'}_{[3,1]}}
{||{\cal S}_{[3,1]}||^2}
+ \frac{ \tilde{\cal S}_{[2,2]}^\rho}{||{\cal S}_{[2,2]}||^2}
+  {\frac{ \tilde{\cal S}_{_\Yup}  }{||{\cal S}_{_\Yup}||^2}}
&
\frac{{\cal S}_{[2,1]}^\rho\cdot {\cal S}_{[1]}}{||{\cal S}_{[2,1]}||^2}
= \frac{\tilde{\cal S}_{[3,1]}^{\rho'\rho''}+\tilde{\cal
S}^{\rho''\rho'}_{[3,1]}}{\frac{27}{4}}
+ \frac{\tilde{\cal S}_{[2,2]}^\rho}{\frac{9}{4}}
+ \frac{\tilde{\cal S}_{_\Yup}  }{\frac{27}{8}}
\nn \\ & \\
\frac{{\cal S}_{[2]}^{\rho'}\cdot {\cal S}_{[2]}^{\rho''} }{||{\cal S}_{[2]}||^4}
\ \stackrel{?}{=}\ \frac{ \tilde{\cal S}_{[3,1]}^{\rho'\rho''}
+\tilde {\cal S}^{\rho''\rho'}_{[3,1]}}
{||{\cal S}_{[3,1]}||^2}
+ \frac{ \tilde{\cal S}_{_\Yup}  }{||{\cal S}_{_\Yup}||^2}
&
\frac{{\cal S}_{[2]}^{\rho'}\cdot {\cal S}_{[2]}^{\rho''} }{||{\cal S}_{[2]}||^4}
= \frac{\tilde{\cal S}_{[3,1]}^{\rho'\rho''}+\tilde{\cal
S}^{\rho''\rho'}_{[3,1]}}{\frac{27}{4}}
+\boxed{\frac{ (\vec\alpha_2^{\rho}\vec p_2)^2 }{18}}
+ \frac{\tilde{\cal S}_{_\Yup}  }{\frac{27}{8}}
\nn \\ & \\
 \frac{({\cal S}_{[2]}^\rho)^2}{||{\cal S}_{[2]}||^4}
\ \stackrel{?}{=}\  \boxed{\frac{ \tilde{\cal S}_{[4]}^\rho }{||{\cal S}_{[4]}||^2}}
+\frac{ \tilde{\cal S}_{[3,1]}^{\rho\rho'}+ {\cal S}^{\rho\rho''}_{[3,1]}}
{||{\cal S}_{[3,1]}||^2}
+ \frac{ \tilde{\cal S}_{[2,2]}^{\rho'}
+ \tilde{\cal S}_{[2,2]}^{\rho''}}{||{\cal S}_{[2,2]}||^2}
\ & \
\frac{({\cal S}_{[2]}^\rho)^2}{||{\cal S}_{[2]}||^4}
= \frac{\tilde{\cal S}_{[3,1]}^{\rho\rho'}+\tilde{\cal
S}^{\rho\rho''}_{[3,1]}}{\frac{27}{4}}
+ \frac{\tilde{\cal S}_{[2,2]}^{\rho'}+\tilde{\cal S}_{[2,2]}^{\rho''}}{\frac{9}{4}}
-\boxed{\frac{ (\vec\alpha_2^{\rho'}\!\vec p_2)^2+(\vec\alpha_2^{\rho''}\!\!\vec p_2)^2}{18}}
\\ &
\end{array}
\ee
and
\be
\sum_{\rho\neq\rho'  }\frac{\tilde{\cal S}_{[3,1]}^{\rho\rho'}\{p\}
\tilde{\cal S}_{[3,1]}^{\rho\rho'}\{p'\}}{\frac{27}{4}}
+\sum_{\rho=x,y,z}\frac{\tilde{\cal S}_{[2,2]}^\rho\{p\}
\tilde{\cal S}_{[2,2]}^\rho\{p'\}}{\frac{9}{4}}
+ \frac{\tilde{\cal S}_{_\Yup} \{p\}\tilde{\cal S}_{_\Yup} \{p'\}}{\frac{27}{8}}
+ \boxed{X\{p,p'\}}
= \nn
\ee
\vspace{-0.4cm}
\be
= \frac{(\vec p_3\vec p_3 \,\!\!')(p_1p_1')}{3}
+ \frac{(\vec p_2\vec p_2 \,\!\!')^2}{8}
+\frac{(\vec p_2\vec p_2 \,\!\!')(p_1p_1')^2}{4} +  \frac{(p_1p_1')^4}{24}
\label{preCauchylev4}
\ee
Here
\be
X\{p,p'\} = -\frac{1}{96} \sum_\rho (\vec\alpha_2^\rho\vec p_2)^2
(\vec\alpha_2^\rho\vec p_2 \!\!\,')^2 + \frac{(\vec p_2\vec p_2 \!\!\,')^2}{16}\neq 0
\ee
is not expressed through the ${\cal S}$-functions
and will be discussed in the next section \ref{viol}.
It makes no direct sense to consider orthogonality at this stage --
because it is expected
only when the $\vec p_4$-dependent terms are included.
However, one can wonder what are orthogonality constraints
on  these $p_4$-dependent terms and if they look resolvable.
Such analysis in the future can help to find a  substitute
of (\ref{scaprmod}), which  better reflects the structure of plane,
rather than ordinary partitions.

Coming back to multiplication rules,
the differences between expected and actual formulas are marked by boxes.
The main of them is the absence of any contribution from $S_{[4]}$ --
but, according to the argument in sec.\ref{Cvskew},
this absence in {\it both} multiplication and Cauchy formulas is {\it not} independent.
Thus it is enough to explain it in just one of these cases.
The simplest is the first line in the multiplication list:
there it is sufficient to look only at the terms $p_3p_1$ and $p_1^4$.
The fact is that the ratio of coefficients in front of these structures is exactly the
same in the sum ${\cal S}^{\rho'\rho''}_{[3,1]}+{\cal S}^{\rho''\rho'}_{[3,1]}$
and ${\cal S}^\rho_{[3]}$.
Indeed, in the latter case the ratio is
$\frac{\vec\alpha^\rho_3\vec p_3}{3}\left(\frac{p_1^3}{6}\right)^{-1} =
\frac{2\vec\alpha^\rho_3\vec p_3}{p_1^3}$,
while in the former case it is rather
$\ \frac{(2\vec\alpha_3^\rho+\vec\beta_3^{\rho'}+\vec\beta_3^{\rho''})\vec p_3p_1}{3}
\left(2\frac{p_1^4}{8}\right)^{-1} =
\frac{4(2\vec\alpha_3^\rho+\vec\beta_3^{\rho'}+\vec\beta_3^{\rho''})\vec p_3}{3p_1^3}\ $
 -- but since $\ \vec\beta_3^{\rho'}+\vec\beta_3^{\rho''}=-\frac{1}{2}\vec\alpha_3^\rho\ $
this is actually the same.
At the same time the same ratio for ${\cal S}_{[4]}^\rho$ is four times bigger:
$\frac{\vec\alpha_3^\rho\vec p_3 p_1}{3}\left(\frac{p_1^4}{24}\right)^{-1}
= \frac{8\vec\alpha_3^\rho\vec p_3}{p_1^4}$,
thus already for these two items one has no chances to add ${\cal S}_{[4]}^\rho$
with any non-vanishing coefficient.

Equally interesting can be the emerging {\it additional} terms in the multiplication rule.
We remind that the product of representations $[2]\otimes[1,1] = [3,1]\oplus[2,1,1]$,
i.e. this is the first example when the product does not contain
the intermediate diagram $[2,2]$, which lies between $[2]+[1,1]=[2,1,1]$
and $[2]\cup[1,1]=[3,1]$ in the lexicographical ordering.
This is exactly the situation reflected in (\ref{sumdia}), i.e.
the $[2,2]$ contribution should vanish for Schur and Macdonald functions,
but show up in the generic Kerov case.
In fact, the Kerov function ${\rm Kerov}_{[2,2]}$ appears in the product
${\rm Kerov}_{[2]}\cdot{\rm Kerov}_{[1,1]}$ with a peculiar coefficient
$g_4g_1^5-3g_4g_2^2g_1+2g_4g_3g_1^2+2g_2^3g_1^3-3g_3g_2g_1^4+g_3g_2^3$
which is the simplest combination of $g$-variables,
vanishing at the Macdonald locus (\ref{macloc}).
The fact that a boxed item appears in the product of
the corresponding 3-Schur functions
$(\vec\alpha_2\vec p_2)^2 \in {\cal S}_{[2]}^\rho\cdot{\cal S}_{[2]}^{\rho''}$
can be a signal that they know about the violation  (\ref{sumdia}) of
the representation-product selection rule --
and have a potential of describing the generic situation, including the Kerov functions.

\section{Anomaly in the Cauchy formula
\label{viol}}

Since the true multiplication formulas at level $4$ are different from the expectation,
i.e. do not fully match  the decomposition formulas (\ref{decolev4}),
we should observe the violation of Cauchy formula.
Indeed, this is what immediately observes in (\ref{preCauchylev4}).
This formula does not contain any reference to ${\cal S}_{[4]}$ --
and this is in accordance with the multiplication rule, where this function
also does not appear, thus this {\it not} a violation.
However, instead it  contains an {\it anomalous} term $X\{p,p'\}$,
reflecting the true difference between multiplication and decomposition,
which we now analyze in a little more detail.

Repeating the argument of sec.\ref{Cvskew} we multiply two copies of (\ref{preCauchylev4})
at $(p,p')$ and $(p,p'')$ and use the fact that expressions at the r.h.s. are bilinear
exponentials, thus they can be substituted by the l.h.s. of still another
(\ref{preCauchylev4})
at $(p,p'+p'')$.
This gives:
\be
\!\!\!\!\!\!
\sum_\rho \left(
\frac{{\cal S}_{[3]}^\rho\cdot{\cal S}_{[1]}}
{||{\cal S}_{[3]}||^2}\otimes \Big({\cal S}_{[3]}^\rho\otimes{\cal S}_{[1]}+
{\cal S}_{[1]}\otimes {\cal S}_{[3]}^\rho\Big)
+  \frac{{\cal S}_{[2,1]}^\rho\cdot{\cal S}_{[1]}}
{||{\cal S}_{[2,1]}||^2}\otimes\Big( {\cal S}_{[2,1]}^\rho\otimes {\cal S}_{[1]}
+ {\cal S}_{[1]}\otimes {\cal S}_{[2,1]}^\rho \Big)
+\frac{{\cal S}_{[2]}^\rho\cdot{\cal S}_{[2]}^\rho}
{||{\cal S}_{[2]}||^4}\otimes {\cal S}_{[2]}^\rho\otimes {\cal S}_{[2]}^\rho
\right)
+ \nn
\ee
\vspace{-0.4cm}
\be
\!\!\!\!\!
+ \sum_{\rho< \rho'}
\frac{{\cal S}_{[2]}^\rho\cdot{\cal S}_{[2]}^{\rho'}}
{||{\cal S}_{[2]}||^4}\otimes\Big( {\cal S}_{[2]}^\rho\otimes{\cal S}_{[2]}^{\rho'}
+ {\cal S}_{[2]}^{\rho'}\otimes {\cal S}_{[2]}^{\rho}\Big)
= \sum_{\rho\neq\rho'}
\frac{\tilde {\cal S}_{[3,1]}^{\rho\rho'}\otimes \Delta\tilde {\cal S}_{[3,1]}^{\rho\rho'}}
{\frac{27}{4}}
+ \sum_\rho \frac{\tilde {\cal S}_{[2,2]}^\rho \otimes \Delta \tilde{\cal S}_{[2,2]}^\rho}
{\frac{9}{4}}
+ \frac{\tilde{\cal S}_{_\Yup}\otimes \Delta\tilde{\cal S}_{_\Yup}}{\frac{27}{8}}
+\! \Delta X
\ee
where  $A\otimes B\otimes C$  denotes $A\{p\}\cdot B\{p'\}\cdot C\{p''\}$.
Substituting the products from the "true" table into the l.h.s. we get:
\be
\Delta\tilde {\cal S}_{[3,1]}^{\rho\rho'} =
\Big({\cal S}_{[3]}^\rho\otimes{\cal S}_{[1]}+
{\cal S}_{[1]}\otimes {\cal S}_{[3]}^\rho\Big)
+ \Big({\cal S}_{[2,1]}^{\rho''}\otimes{\cal S}_{[1]}+
{\cal S}_{[1]}\otimes {\cal S}_{[2,1]}^{\rho''}\Big)
+ {\cal S}_{[2]}^{\rho''}\otimes {\cal S}_{[2]}^{\rho''}
+ \Big( {\cal S}_{[2]}^\rho\otimes{\cal S}_{[2]}^{\rho'}
+ {\cal S}_{[2]}^{\rho'}\otimes {\cal S}_{[2]}^{\rho}\Big)
\nn \\
\Delta \tilde{\cal S}_{[2,2]}^\rho =
\Big({\cal S}_{[2,1]}^{\rho}\otimes{\cal S}_{[1]}+
{\cal S}_{[1]}\otimes {\cal S}_{[2,1]}^{\rho}\Big)
+ {\cal S}_{[2]}^{\rho'}\otimes {\cal S}_{[2]}^{\rho'}
+ {\cal S}_{[2]}^{\rho''}\otimes {\cal S}_{[2]}^{\rho''}
\nn \\
\Delta\tilde {\cal S}_{_\Yup} =
\sum_\rho \Big({\cal S}_{[2,1]}^\rho\otimes {\cal S}_{[1]} +
{\cal S}_{[1]}\otimes {\cal S}_{[2,1]}^\rho\Big)
+\sum_{\rho\neq\rho'}
{\cal S}_{[2]}^\rho\otimes{\cal S}_{[2]}^{\rho'} \ \ \ \
\ee
and additionally at the l.h.s. we have a contribution
from the boxed terms in the multiplication formulas:
\be
-\sum_\rho \frac{ (\vec\alpha_2^{\rho}\vec p_2)^2 }{18}
\Big({\cal S}_{[2]}^{\rho'}-{\cal S}_{[2]}^{\rho''}\Big)\otimes
\Big({\cal S}_{[2]}^{\rho'}-{\cal S}_{[2]}^{\rho''}\Big)
= -\frac{1}{72}\sum_\rho (\vec\alpha_2^{\rho}\vec p_2)^2
\Big((\vec\alpha_2^{\rho'}-\vec\alpha_2^{\rho''})\vec p_2 \!\!\,'\Big)
\Big((\vec\alpha_2^{\rho'}-\vec\alpha_2^{\rho''})\vec p_2 \!\!\,''\Big) \neq 0
\ee
This is exactly the same as the $\Delta X$ term at the r.h.s:
\be
\Delta X:= X\{p,p'+p''\}-X\{p,p'\}-X\{p,p''\}
= \frac{2}{16}(\vec p_2\vec p_2 \!\!\,')(\vec p_2\vec p_2 \!\!\,'')
-\frac{2}{96}\sum_\rho (\vec\alpha_2^\rho\vec p_2)^2(\vec\alpha_2^{\rho}\vec p_2 \!\!\,')
(\vec\alpha_2^{\rho}\vec p_2 \!\!\,'')
\ee
Thus the anomaly in Cauchy summation formula can indeed be used to
measure the deviation of multiplication from skew decomposition.

\section{Conclusion}

In this note we explained the nearly rigid relation between Cauchy summation formula
(\ref{Cauchy})
and the equivalence of the structure constants in multiplication and skew-decomposition
formulas (\ref{decoskew}) and (\ref{ringdu}):
\be
\boxed{
\sum_\sigma {\bf S}_\sigma\{p\}S_\sigma\{p'\} = \exp \Big({\bf bilinear}(p,p')\Big)
\ \ \Longleftrightarrow \ \ \ {\bf N}={\bf M}
}
\ \ \ {\rm for}\ \
\begin{array}{c}
{\bf S}_\sigma = \sum_{\sigma',\sigma''} {\bf N}^\sigma_{\sigma'\sigma''}
{\bf S}_{\sigma'}\cdot{\bf S}_{\sigma''} \\
S_\sigma\{p'+p''\} = \sum_{\sigma',\sigma''}
{\bf M}^\sigma_{\sigma'\sigma''} S_{\sigma'}\{p'\}S_{\sigma''}\{p''\}
\end{array}
\nn
\ee
We illustrated this fact by an important example of the would-be 3-Schur functions
for 4-box plane partitions:
mismatches  are simultaneously present and well correlated
in expressions of both kinds.
Thus it is sufficient to cure just one of the anomalies --
the other will be automatically fixed.
This, however, remains to be done.
More generally,
this paper can help to understand the abundance of Cauchy formula,
i.e. why it appears in one and the same form for a broad variety of special functions
and why it is  actually {\it not} restricted to the case of {\it Young} diagrams
in the role of $\sigma$ and {\it one}-parameter sets of time-variables in the role of $p$.

\section*{Acknowledgements}

My work is partly supported by the grant of the
Foundation for the Advancement of Theoretical Physics BASIS,
by RFBR grant 19-02-00815
and by the joint grants 17-51-50051-YaF, 18-51-05015-Arm,
18-51-45010-Ind,  RFBR-GFEN 19-51-53014.
I also acknowledge the hospitality of KITP and
partial support by the National Science Foundation
under Grant No. NSF PHY-1748958
at the final stage of this project.

\end{document}